\documentstyle[aps]{revtex}

\begin{document}
\input{epsf}

\draft

\title{Coherent Disintegration and Stability of 
Vortices in Trapped Bose Condensates}


\author{H. Pu$^{1,2}$, C. K. Law$^{2}$, J. H. Eberly$^{2}$
and N. P. Bigelow$^{1}$}

\address{$^1$Department of Physics and Astronomy, and
Laboratory for Laser Energetics\\
the University of Rochester, Rochester, NY 14627\\ $^2$Rochester
Theory
Center for Optical Science and Engineering,
and Department of Physics and Astronomy\\ the University of
Rochester, Rochester, NY 14627}

\date{\today}

\maketitle

\begin{abstract}
We consider the intrinsic stability of the vortex states of a
pure Bose-Einstein condensate confined in a harmonic potential
under the effects of coherent atom-atom interaction. We
find that stable vortices can be supported, and 
that vortex stability can be controlled by 
changing the inter-particle
interaction strength.
At unstable regimes, a vortex
will spontaneously disintegrate into states with different angular
momenta even without external 
perturbations, with the lifetime determined by its imaginary
excitation frequencies.
\end{abstract}


\section{Introduction}
Vortices and their motions have long been an important branch of
fluid mechanics. With the discovery of superfluid helium II, a new
idea developed: that the circulation in a superfluid vortex must be
quantized\cite{1}. The consequences of these quantized vortices is
profound and the understanding of vortex dynamics plays a key role
in the current understanding of superfluidity. Moreover, the
detection of individual singly quantized vortices has vividly
established the true macroscopic quantum nature of these remarkable
degenerate fluids.  Intimately related to the observation of the
superfluid state of $^4$He is the evidence for the concurrent
formation of a Bose-Einstein condensate (BEC)\cite{bec}. An
important challenge is to clarify the link between these fundamental
and important phenomena --- Bose-Einstein condensation,
superfluidity and the formation of a macroscopic quantum state.

The recent observation of Bose-Einstein condensation in dilute alkali
vapors\cite{2,3,3.4} has presented a striking new system for
investigation; that of the dilute degenerate Bose $gas$. The alkali
BECs differ fundamentally from the helium BEC in several crucial
ways. BEC in both bulk liquid helium and the dilute helium ``gas"
are free systems(the ``gas'' BEC is
created by introducing helium into a porous glass known as
Vycor\cite{reppy}).
By contrast, the alkali vapor BECs, although free
of container walls (and/or the Vycor host), are created within
the confines of a trapping potential. There is another major
difference: in the trapped alkali condensates, samples can be
prepared in which essentially $all$ of the atoms are Bose
condensed. By contrast, in bulk superfluid $^4$He, although the
superfluid fraction can be near unity, momentum distribution
measurements have shown that the bulk condensate fraction is closer
to 0.1 with the remainder of the particles in finite momentum
states. As researchers improve their ability to create and
manipulate these new trapped gaseous condensates, a series of
important questions naturally arise: Does the gaseous BEC support
superflow? Is it indeed a superfluid? Are there stable vortices? This
last question is the subject of this paper.

The problem of vortex state excitations has been recently treated by
others. Sinha\cite{sin} investigated the low-lying modes under
Thomas-Fermi limit and
Dodd {\em et al.}\cite{6}
obtained the normal mode 
spectrum of a single quantized vortex state as a function of the number
of condensate atoms for a BEC confined in a TOP trap\cite{2}.
However, the important question of vortex stability was not addressed
by these authors. More recently, Rokhsar\cite{7} studied the stability
properties of the trapped vortices and argued that vortices are
unstable due to the existence of a bound state inside the vortex core.
However, throughout his analysis,
the transition from a vortex state to a core state requires the presence
of thermal atoms which serve as a reservoir to conserve the energy
and angular momentum in the process. Hence, the more fundamental
question concerning the intrinsic stability of an isolated vortex
(i.e., without the external influence of thermal atoms) remains
unanswered. 

In this paper, we approach the problem by assuming
that all atoms are in the
condensate such that scattering with thermal background atoms can be
neglected.  This allows us to focus on the {\em
intrinsic coupling} within and between different vortex states and on
the effect of this coupling on vortex stability.
(Here, we use the word ``intrinsic'' to emphasize coherent
coupling between the condensate atoms.) 
We find that stable vortex states can in fact be supported and show
that whether a vortex state is stable or not is determined by its angular
momentum and the nonlinear inter-particle interaction strength.
Furthermore, we point out that the lifetime of an unstable vortex can
be directly determined from the frequencies of the collective excitations.

The paper is organized as follows.
In Section II, we discribe our physical model and define the stability 
criterion. Our main results are presented in Section III, where
the stable and unstable regions of trapped vortices are identified.
We also present a physical interpretation of the meaning of the
instability. Finally, we give a summary and compare our work with
others in Section IV.

\section{Physical Model}
To simplify our calculations, we consider a condensate confined in
a 2d isotropic harmonic potential with trap frequency $\omega_0$
at zero temperature. 
In current experiments, condensates are
achieved in 3d traps with cylindrical symmetry. A quasi-2d situation
can be realized when $\omega_{\bot} \ll
\omega_z$, where $\omega_{\bot}$ and
$\omega_z$ are transverse and longitudinal trap frequencies
respectively\cite{2d3d}. In this limit, one can produce a 
``pancake''-shaped condensate with
all the atoms lying in the lowest harmonic oscillator state in
$z$-direction and hence,
the degree of freedom in
$z$-coordinate is frozen. 
Our treatment lies within the Hartree-Fock-Bogoliubov approximation.
First, we calculate the macroscopic
wavefunctions of the condensate in
a vortex state. Next, we find the
collective excitation frequency $\omega$ of the state.
The stability criterion is defined
as Im($\omega$)=0\cite{4}, selected because if
Im$(\omega ) \ne 0$, then fluctuations around the condensate can grow
exponentially in time and hence induce instability.

In the Gross-Pitaevskii treatment\cite{8}, the energy for $N$ condensed
bosons
of mass $m$ is given by the functional:
\begin{equation}
\frac{E(\Psi_{\kappa})}{N}=\int d{\bf r}\,\,(\Psi_{\kappa}^*
\hat{T}\Psi_{\kappa}+{\hat V}
|\Psi_{\kappa}|^2+\frac{1}{2}NU|
\Psi_{\kappa}|^4)\,\, ,
\end{equation}
Here \begin{equation}
\Psi_{\kappa}({\bf r})=\Phi_{\kappa}(r)e^{i\kappa \theta}\, ,
\,\kappa=0,\pm 1,\pm 2,...
\end{equation}
represents the wavefunction of the macroscopic
vortex state with
azimuthal angular momentum $\kappa \hbar$.
${\hat T}=-\hbar^2 \nabla^2/2m,
{\hat V}=m \omega_0^2r^2/2$
are the kinetic and potential energy operators, respectively, and the
coupling constant
$U$ describes the interactions
between condensate atoms.
In the quasi-2d situation considered here,
the coupling constant takes the form 
$U=4\sqrt{\pi} \hbar \omega_z \xi_z a$\cite{14}, where
$\xi_z=\sqrt{\hbar/2m\omega_z}$ is the harmonic oscillator length in
$z$-dimension.
In our analysis, the
solution that minimizes Eq.~(1) is found iteratively using a finite
elements method (FEM)\cite{9,10}. In our calculations,
we normally used 20 elements, with 2 nodes and 3 degrees of
freedom for each element. This numerical method is very efficient,
and it typically
took no more than a few minutes
to find the wavefunction $\Psi_{\kappa}$ in a Cray-YMP2E/232.

With the solution of $\Psi_{\kappa}$ at our disposal, we can now
calculate collective excitation frequencies by solving Bogoliubov
equations\cite{6}: \begin{mathletters}
\begin{eqnarray}
({\cal L}-\hbar \omega_{\lambda}-\mu_{\kappa}) \,u_{\lambda}({\bf
r})
+NU[\Psi_{\kappa}({\bf r})]^2\,v^*_{\lambda}({\bf r}) & = & 0,\\
NU[\Psi_{\kappa}^*({\bf r})]^2\,u_{\lambda}({\bf r}) + ({\cal
L}+\hbar \omega_{\lambda}-\mu_{\kappa}) \,v^*_{\lambda}({\bf
r}) & = & 0,
\end{eqnarray}
\end{mathletters}
\noindent where $\mu_{\kappa}$ is the chemical potential for state
$\Psi_{\kappa}({\bf r})$,
$u_{\lambda}({\bf r}), v_{\lambda}({\bf r})$ are normal mode
functions with mode frequency $\omega_{\lambda}$, and ${\cal
L}=\hat{T}+\hat{V}+ 2NU|\Psi_{\kappa}({\bf r})|^2$.
It is straightforward to
show that if $\Psi_{\kappa}({\bf r})$ is given by Eq.~(2), then
$u_{\lambda}({\bf r}), v_{\lambda}({\bf r})$ must have definite
angular momentum compositions $\kappa_u \hbar$ and $\kappa_v
\hbar$ respectively such that $u_{\lambda}({\bf
r})=\tilde{u}_{\lambda}(r) e^{i\kappa_u \theta}, v_{\lambda}({\bf
r})=\tilde{v}_ {\lambda}(r)e^{i\kappa_v \theta}$,
and $\kappa_u + \kappa_v = 2\kappa$\cite{6}.

\section{Results and Interpretation}
Eqs.~(3) were transformed to an eigenvalue problem for a finite-sized
matrix and solved using the FEM.
Our goal
here is to find mode frequencies with non-zero imaginary part, in
order to determine the vortex stability. As in the case of ground state,
the vortex stability properties for a condensate 
with repulsive inter-particle interaction
are drastically different from that for a condensate
with attractive interaction. We will discuss these two cases separately.

{\em Repulsive interaction, i.e., $U>0$} --- When we calculate the
collective excitation frequencies of a single quantized ($\kappa = 1$)
vortex state $\Psi_1$, we find that all the excitation frequencies are
real, which means that $\Psi_1$ is {\em always stable}.
Next, we consider a
double quantized ($\kappa = 2$) vortex state $\Psi_2$.
Here we find that
complex frequencies only exist for $\kappa_u=0$ and $\kappa_v=4$
(Without loss of generality, we assume $\kappa>0$, and
$\kappa_v>\kappa>\kappa_u$.).  We find that for any other pairs of
$(\kappa_u,\kappa_v)$, the excitation frequencies are all real.
Furthermore, for
values of $NU$ for which the vortex is unstable, we find that there
exists at most one complex frequency. Fig.~1(a) shows the imaginary
part of the complex frequency as a function of interaction strength
$NU$. As we can see,
in this particular channel [i.e. choice of ($\kappa_u,\kappa_v$)],
the parameter space of $NU$ is divided into alternating stable and
unstable regions.  In the unstable regions, the inverse of Im($\omega$)
determines the lifetime of the unstable vortex. For
the parameter range described in Fig.~1, the most unstable vortex state
will decay after several periods of the harmonic trapping potential. We
stress that the details of how the condensate will evolve under these
instabilities is beyond the capability of the mean-field treatment and
requires further investigation.

\begin{figure}

\centerline{
\epsfxsize=3.5in
\epsfbox{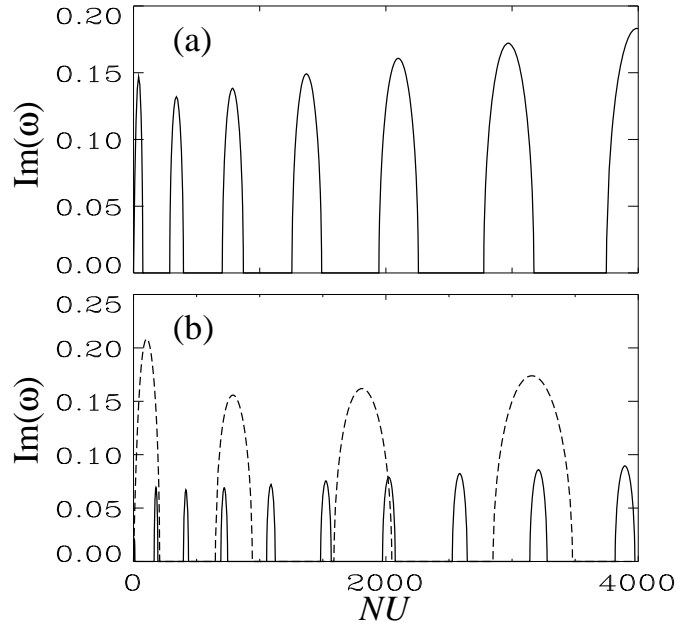}}
\caption{(a) Imaginary part of the complex frequency of a double
quantized vortex state $\Psi_2$ as a function of interaction strength
$NU$, for $\kappa_u=0$ and $\kappa_v=4$. (b) Same as (a) for a
triple quantized vortex state $\Psi_3$.
Solid line: $\kappa_u=0$ and
$\kappa_v=6$; dashed line: $\kappa_u=1$ and $\kappa_v=5$.
Frequency is in units of trap frequency $\omega_0$, $U$ in units of
($\hbar \omega_0 \xi^2$),
where $\xi =(\hbar /2m \omega_0)^{1/2} $ is the harmonic oscillator
length.}
\end{figure}

For a general
state $\Psi_{\kappa}$, our numerical calculations show that there are
($\kappa-1$) unstable channels that possess complex excitation
frequencies; those with $\kappa_u=0,1,...,\kappa-2$ and $\kappa_v=
2\kappa-\kappa_u$.
Fig.~1(b) shows the imaginary part of
the complex frequency for a triple quantized vortex state $\Psi_3$. We
can see a similar pattern as in Fig.~1(a), but here, there are two
unstable channels. Each channel shows its own quasi-periodic
behavior as a function of $NU$. The two channels have quite
different ``period'' and characteristic width of unstable regions. At
first look, this may
appear rather unexpected.  To interpret this behavior
we will show that
each unstable region in $NU$-space
represents a decay channel in which
two atoms from the given vortex state scatter into two new states, with
angular momenta $\kappa_u \hbar$ and $\kappa_v \hbar$,
respectively, thus inducing instability in that initial vortex state
$\Psi_{\kappa}$.

First, let us define a boson field operator as: $\hat{\Psi}({\bf r})
\equiv \sqrt{N} \Psi_{\kappa}
({\bf r}) +\hat{\psi}({\bf r})$, where the 
$c$-number $\Psi_{\kappa}({\bf r})$ 
denotes the one-body wavefunction for the
condensate and $\hat{\psi}({\bf r})$ is the field operator for
the fluctuation part\cite{11}. 
The second quantized Bogoliubov Hamiltonian reads:
\begin{eqnarray*}
\hat{K}_B=\int_{}^{} d{\bf r} \,\,\hat \psi^{\dagger} ({\bf r})
[{\cal L}-\mu_{\kappa}]\hat{\psi} ({\bf r})+[\frac{1}{2}NU
\int_{}^{} d{\bf r} \,\,\hat{\psi}^{\dagger} ({\bf r})
\hat{\psi}^{\dagger} ({\bf r}) \Psi_{\kappa} \Psi_{\kappa}
+h.c.]\, ,
\end{eqnarray*}
where the $c$-number part independent of $\hat \psi ({\bf r})$ 
has been neglected.
We can further decompose
$\hat{\psi}({\bf r})$ as $\hat \psi ({\bf r})
=\sum_{n,\alpha}a_{n,\alpha} \phi_{n,\alpha} ({\bf r})$, where
$a_{n,\alpha}$ is an annihilation operator associated with a
single-particle state $\phi_{n,\alpha}$. The set of states
$\left\{ {\phi_{n,\alpha}} \right\}$ is defined as the eigenvectors of
$\cal{L}$ with eigenvalues $\epsilon_{n,\alpha}$, i.e., $\cal
{L}$$\phi_{n,\alpha}=\epsilon_{n,\alpha} \phi_{n,\alpha}$, with
subscripts $(n,\alpha)$ labeling the radial and angular quantum
number respectively. Hamiltonian $\hat K_B$ may then be rewritten as:
$\hat{K}_B=\hat H_0 + \hat H_I$, where
\begin{eqnarray*}
\hat{H}_0 & = & \sum_{n,\alpha}
(\epsilon_{n,\alpha}-\mu_{\kappa})a_{n,\alpha}^{\dagger}
a_{n,\alpha},\\
\hat{H}_B & = & \sum_{n_u,\kappa_u} \sum_{n_v,\kappa_v} \Lambda
\left( {n_u,\kappa _u;n_v,\kappa _v}
\right)
a_{n_u,\kappa _u}^{\dagger} a_{n_v,\kappa
_v}^{\dagger}+h.c., 
\end{eqnarray*}
and
\begin{equation}
\Lambda \left( {n_u,\kappa _u;n_v,\kappa _v} \right)=
\frac{1}{2}NU \int_{}^{} d{\bf r} \,\,\phi^*_{n_u,\kappa_u} ({\bf r})
\phi^*_{n_v,\kappa_v}({\bf r}) \Psi_{\kappa}
({\bf r}) \Psi_{\kappa}({\bf r})\,\,.
\end{equation}
In the interaction picture,
$a_{n,\alpha}^{\dagger}(t)=a_{n,\alpha}^{\dagger}e^{i(
\epsilon_{n,\alpha}-\mu_{\kappa})t}$, and the Hamiltonian is
given by:
\begin{equation}
\hat H_I (t)= \sum_{n_u,\kappa_u} \sum_{n_v,\kappa_v} \Lambda
\left( {n_u,\kappa _u;n_v,\kappa _v}
\right)e^{i(\epsilon_{n_u,\kappa_u}+\epsilon_{n_v,\kappa_v} -
2\mu_{\kappa}) t} a_{n_u,\kappa _u}^{\dagger} a_{n_v,\kappa
_v}^{\dagger}+h.c.\,
\label{Happrox}
\end{equation}
$\Lambda \left( {n_u,\kappa _u;n_v,\kappa _v} \right)$ is non-zero         
only when $\kappa_u + \kappa_v=2\kappa$, which is a direct
consequence of the conservation of angular momentum. The
interaction described by
Hamiltonian (\ref{Happrox}) is analogous to parametric processes in
quantum optics where instability can occur under certain
conditions. For example, one can build up large numbers of photon
pairs (signal and idle) from the vacuum via parametric down
conversion if the field frequencies satisfy a parametric resonance
condition. In our case, the fluctuation in mode pair
($\phi_{n_u,\kappa_u},\phi_{n_v,\kappa_v}$) grows exponentially
when
\begin{equation}
|\epsilon_{n_u,\kappa_u}+\epsilon_{n_v,\kappa_v} -
2\mu_{\kappa} |< \Lambda \left
( {n_u,\kappa _u;n_v,\kappa _v} \right), \label{condition}
\end{equation}
and hence the vortex is unstable under such resonance condition. We
emphasize that the instability implied in this picture is purely quantum
mechanical. The atoms in the vortex can {\em spontaneously disintegrate}
into $\phi _{n_u,\kappa _u}$ and $\phi _{n_v,\kappa _v}$ states
without the need of external (classical) perturbations, such as the     
interaction with the thermal background gases or perturbation of the
trap.

For a $\kappa =1$ vortex, our numerical calculations show that there
exists {\em no} particle states that 
satisfy the resonance condition (\ref{condition}), in
support of our prediction that a single quantized vortex state is always
stable for $U>0$. For a $\kappa =2$ vortex, we find a pair of particle
states ($\phi_{0,0},\phi_{0,4}$) indeed satisfy inequality
(\ref{condition}). In the weak coupling limit, we can calculate the
decay rate of the double quantized vortex state $\Psi_2$ using the
Hamiltonian (\ref{Happrox}) by neglecting all the nonresonant terms
(i.e., keeping only terms with $n_u=n_v=0, \kappa_u=0,
\kappa_v=4$ ). The results are shown in Fig.~2 along with the
imaginary part of the complex excitation frequencies of vortex state
$\Psi_2$. We can see a clear
qualitative agreement between the two results. The agreement can be
significantly improved if the contribution from states
($\phi_{1,0},\phi_{1,4}$) is also included in calculating the decay
rate (see Fig.~2).
We remark that although useful for interpreting our vortex stability
predictions, the parametric resonance picture is valid only for the weak
interaction regime. This is because a strong interaction can drastically
change the frequencies of the oscillators and introduce mixing among
different particle states. Further work would be necessary in order to
understand all aspects of the complex structure shown in Fig.~1,
particularly for large $NU$.

\begin{figure}
\centerline{
\epsfxsize=3.5in
\epsfbox{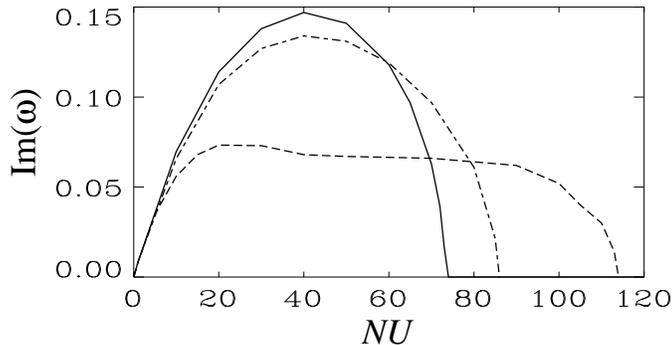}}
\caption{Solid line: imaginary part of the complex frequency of a
double quantized vortex state $\Psi_2$ as a function of interaction
strength $NU$
for $\kappa_u=0$ and $\kappa_v=4$; dashed line: decay rate of the
vortex state $\Psi_2$ as a function of $NU$ calculated using
Hamiltonian (4), only contributions from the resonant states
($\phi_{0,0},\phi_{0,4}$) are included; dot dashed line: same as
dashed line, but include contributions from two more states
($\phi_{1,0},\phi_{1,4}$).} \end{figure} 

{\em Attractive interaction, i.e., $U<0$} --- A condensate with strong
attractive inter-particle interaction
is known to be subject to collapse. However, a metastable condensate
with a small
number of atoms can still exist\cite{3,12}. Fig.~3 shows the imaginary
part of the complex excitation frequency for a single 
and a double quantized vortex
state as functions of $NU$. 
Fig.~3(a) shows that $\Psi_1$ is stable for sufficiently
small attractive interaction, but unstable for larger 
interaction strength. For $\Psi_2$, as we can see from Fig.~3(b),
the channel
($\kappa_u=0,\kappa_v=4$) possesses complex frequency for all
negative values of $NU$ instead of showing a quasi-periodic pattern
as in the case of repulsive interaction.
Furthermore, we find that, similar to $\Psi_1$,
other channels which are stable for
$NU>0$ become consistently unstable for sufficiently large 
$|NU|$ [We only show two such channels
in Fig.~3(b).]. Our calculations show that, for $U<0$,
stable vortices only exist for a single quantized vortex
state in the weak interaction regime [see Fig.~3(a)]; multiple
quantized vortex state (i.e., $\kappa>1$) is always 
unstable.
It has been speculated that the existence of vortices
may help stabilize a condensate with 
negative scattering length\cite{13.5}.
However, as we show here, although such vortices may seem to be
more stable against the collapse when compared to the ground state,
they remain fundamentally unstable and small fluctuations will
eventually destroy such vortices.

\begin{figure}
\centerline{
\epsfxsize=3.in
\epsfbox{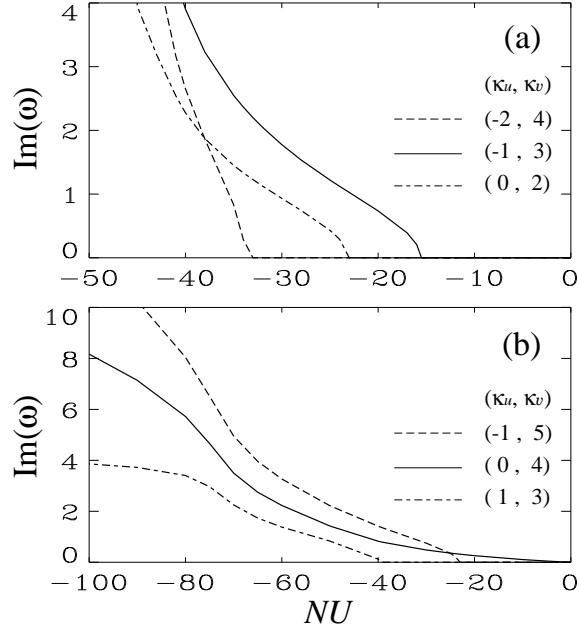}}
\caption{Imaginary part of the complex frequency of a single (a)
and a double (b)
quantized vortex state as a function of interaction strength
$NU$, with $NU<0$.}
\end{figure} 

\section{Summary and Discussion}
In summary, we have calculated the collective excitation frequencies
of a Bose-Einstein condensate 
in a vortex state and have established intrinsic stability regions
for these vortices. We have shown that, even without any perturbation,
an unstable vortex can still decay spontaneously.
For repulsive inter-particle
interaction, we found that single quantized vortices are 
always stable, while 
imaginary excitation modes divide the
interaction energy axis ($NU$) 
of multiple quantized vortices
($\kappa>1$)
into alternating stable
and unstable regions. Hence, one can control
the vortex stability by varying the value of interaction
strength, which in turn can be achieved by changing the scattering
length\cite{13,nat}, particle number or trap frequency. 
This provides us with the possibility of studying condensate
evolution under the effect of imaginary modes.

For a condensate in vortex state,
there may exist quasiparticle states with negative frequencies.
One such negative frequency state
was identified by Dodd {\em et al.}
in Ref.\cite{6}. The presence of negative frequencies implies that
there exist states with lower energy. However,
this does not necessarily mean that the condensate is
unstable if no mechanism exists to drive the system
to these lower energy states\cite{sta}.
In Ref.\cite{7}, Rokhsar considered the instability arising from the
incoherent interactions between condensate and thermal atoms, which
induce the transition to the negative frequency core state.
In contrast, 
in the present paper, we study the {\em intrinsic
stability of vortices in a pure condensate}, excluding such incoherent
processes while focusing on the coherent interactions within the
condensate.
In our work, 
instability occurs as a {\em coherent process}
such that an unstable vortex state will
disintegrate into different angular momenta states.
We found that 
stable vortices {\em can} be supported
in harmonic traps as long as the 
temperature is low enough such that 
the effects of thermal atoms are insignificant.
At temperatures when
thermal atoms cannot be neglected, both coherent and incoherent
processes will be present and each will have its effect on 
vortex stability.
It remains to be seen which process will be dominant.
Further investigations should also include the possible                 
influence of trap anisotropy and the dynamics of the disintegration
processes.

Recently, vortex stability in 2d harmonic trap
is studied by Caradoc-Davies {\em et al.} 
through a direct numerical simulation\cite{simu}. 
In that study, a blue detuned laser beam is applied to perturb
the condensate in a vortex state.
They found that the single quantized vortex is indeed stable, while
a double quantized vortex can disintegrate into unit
vortices under external perturbation. These results are
consistent with ours presented in this paper.

Finally, as an example, 
let us consider a $^{23}$Na condensate (scattering
length $a \approx 3$ nm) in a harmonic trap with
$\omega_{\bot}=2\pi \times 10$ Hz and
$\omega_z=2\pi \times 200$ Hz, in units of $\hbar \omega_{\bot}
\xi_{\bot}^2$, $U\approx 0.02$. The plotted range of $NU$ from 0
to 4000 in Fig.~1 corresponds to particle number ranging
from 0 to $2 \times 10^5$, well within the capability of current
experiments. Recently, several methods on how to generate vortex
states in alkali atomic BECs have been proposed\cite{15}. With
current technology and fast progress on this field, our study on vortex
stability should be
experimentally testable in the near future.

\acknowledgements
This research was supported by NSF grants PHY-9415583 and PHY-9457897,
and the David and Lucile Packard Foundation. NPB thanks Prof. D.
S. Rokhsar for comments regarding Ref.~(9).



\begin{references}


\bibitem{1} L. Onsager, Nuovo Cimento Suppl. {\bf 6}, 249 (1949);
R. P. Feynman, in {\em Progress in Low Temperature
Physics I}, edited by C. J. Gorter, North-Holland Publishing
Co.,1955; W. F. Vinen, Proc. Roy. Soc. {\bf A260},
218 (1961). 
\bibitem{bec} F. London, Nature {\bf 141}, 643 (1938).
\bibitem{2} M. H. Anderson {\em et al.}, Science {\bf 269}, 198
(1995). 
\bibitem{3} C. C. Bradley {\em et al.}, Phys. Rev. Lett. {\bf 75}, 1687
(1995). 
\bibitem{3.4} K. B. Davis {\em et al.}, Phys. Rev. Lett. {\bf 75}, 3969
(1995).
\bibitem{reppy} B. C. Crooker, B. Hebral, E. N. Smith, Y. Takano,
and J. D. Reppy, Phys. Rev. Lett. {\bf 51}, 666 (1983).
\bibitem{sin} S. Sinha, Phys. Rev. A {\bf 55}, 4325 (1997).
\bibitem{6} R. J. Dodd {\em et al.}, Phys. Rev. A {\bf 56}, 587
(1997).
\bibitem{7} D. S. Rokhsar, Phys. Rev. Lett. {\bf 79}, 2164 (1997).
\bibitem{2d3d} We undertook a calculation of the excitation spectrum of 
a single quantized vortex state confined in a 2d trap, and compared with
the results shown in Ref.\cite{6}. We found that our results are in 
quantitative agreement with theirs which are obtained for a 3d trap
with $\omega_z / \omega_{\bot} \approx 2.3$.
\bibitem{4} W. I. Glaberson, W. W. Johnson and R. M.
Ostermeier, Phys. Rev. Lett. {\bf 33}, 1197 (1974).
\bibitem{8} L. P. Pitaevskii, Sov. Phys. JETP {\bf 13}, 451 (1961);
E. P. Gross, Nuovo Cimento {\bf 20}, 454 (1961). 
\bibitem{14} Tin-Lun Ho and Michael Ma, e-print
cond-mat/9703034.
\bibitem{9} B.
D. Esry, Phys. Rev. A {\bf 55}, 1147 (1997). 
\bibitem{10} L. R.
Ram-Mohan {\em et al.}, Computers in Phys. {\bf Jan/Feb}, 50
(1990); K. -J. Bathe, {\em Finite Element Procedures in Engineering
Analysis}, Prentice-Hall, Englewood Cliffs, NJ, 1982.
\bibitem{11} Alexander L. Fetter, Annals of Phys. {\bf 70}, 67
(1972).
\bibitem{12} M. Houbiers and H. T. C. Stoof, Phys. Rev. A {\bf
54}, 5055 (1996); Hualin Shi and Wei-Mou Zheng, Phys. Rev. A
{\bf 55}, 2930 (1997); T. Bergeman, Phys. Rev. A {\bf 55}, 3658
(1997).
\bibitem{13.5} F. Dalfovo and S. Stringari, Phys. Rev. A {\bf 53},
2477 (1996).
\bibitem{13} E. Tiesinga {\em et al.}, Phys. Rev. A {\bf 47}, 4114
(1993); P. O. Fedichev {\em et al.}, Phys. Rev. Lett. {\bf 77}, 2913
(1996).
\bibitem{nat} S. Inouye {\em et al.}, Nature {\bf 392}, 151 (1998).
\bibitem{sta} A detailed analysis of the relationship between
normal mode frequencies and condensate stability is reported in:
H. Pu, and N. P. Bigelow, submitted. 
\bibitem{simu} B. M. Caradoc-Davies, R. J. Ballagh, K. Burnett,
and P. B. Blakie, submitted. An outline of this paper can be found
in the web: http://www.vuw.ac.nz/~gardiner/otago/vortexpaper/vortexpaper.html
\bibitem{15} Karl-Peter Marzlin {\em et al.}, Phys. 
Rev. Lett. {\bf 79}, 4728 (1997);
R. Dum {\em et al.}, Phys. Rev. Lett. {\bf 80}, 2972 (1998);
Karl-Peter Marzlin and Weiping Zhang, Phys. Rev. A {\bf 57},
3801 (1998); Eric L. Bolda and Dan F. Walls, Report No.
cond-mat/9708189.
\end{references}
\end{document}